\documentclass{article}
\usepackage[utf8]{inputenc}
\usepackage[T1]{fontenc}
\usepackage{geometry}
\usepackage{amsfonts}
\usepackage{spconf}
\usepackage{amsmath}
\usepackage{graphicx}
\usepackage{array}
\usepackage{float}
\usepackage[space]{grffile}
\usepackage{stmaryrd}
\usepackage{subfig}
\usepackage[section]{placeins}
\usepackage{algpseudocode}
\usepackage{algorithm}
\usepackage{color}
\usepackage{enumitem}
\usepackage{times}
\usepackage[makeroom]{cancel}
\usepackage{MnSymbol}
\usepackage{longtable}
\usepackage{url}

\usepackage{etoolbox}
\apptocmd{\thebibliography}{\footnotesize}{}{}
\usepackage{balance}

\newcommand{\E}{\mathbb{E}}

\def\mbf#1{\mathbf{#1}}
\def\mbb#1{\mathbb{#1}}
\def\bs#1{\boldsymbol{#1}}

\def\myparagraph#1{\smallskip\smallskip\noindent\textbf{#1. \hspace{.1cm}}}

\algrenewcommand\algorithmicforall{\textbf{independently for all}}

\makeatletter
\newcounter{savesection}
\newcounter{apdxsection}
\renewcommand\appendix{\par
  \setcounter{savesection}{\value{section}}%
  \setcounter{section}{\value{apdxsection}}%
  \setcounter{subsection}{0}%
  \gdef\thesection{\@Alph\c@section}}
\newcommand\unappendix{\par
  \setcounter{apdxsection}{\value{section}}%
  \setcounter{section}{\value{savesection}}%
  \setcounter{subsection}{0}%
  \gdef\thesection{\@arabic\c@section}}
\makeatother

\setlist[itemize]{label=$\triangleright$}

\usepackage[hang,flushmargin]{footmisc}

\title{Speech enhancement with variational autoencoders \\ and alpha-stable distributions}
\name{Simon Leglaive\textsuperscript{\normalfont 1}, Umut \c Sim\c sekli\textsuperscript{\normalfont 2}, Antoine Liutkus\textsuperscript{\normalfont 3}, Laurent Girin\textsuperscript{\normalfont 1,4}, Radu Horaud\textsuperscript{\normalfont 1} \thanks{This work is supported by the ERC Advanced Grant VHIA \#340113.}}
\address{
\textsuperscript{1}Inria Grenoble Rh\^one-Alpes, France, \hspace{10pt}
\textsuperscript{2}LTCI, T\'{e}l\'{e}com ParisTech, Université Paris-Saclay, France\\
\textsuperscript{3}Inria and LIRMM, France, \hspace{10pt}
\textsuperscript{4}Univ. Grenoble Alpes, Grenoble INP, GIPSA-lab, France
}

\begin{document}
\ninept
\maketitle

\begin{abstract}
This paper focuses on single-channel semi-supervised speech enhancement. We learn a speaker-independent deep generative speech model using the framework of variational autoencoders. The noise model remains unsupervised because we do not assume prior knowledge of the noisy recording environment. In this context, our contribution is to propose a noise model based on alpha-stable distributions, instead of the more conventional Gaussian non-negative matrix factorization approach found in previous studies.
We develop a Monte Carlo expectation-maximization algorithm for estimating the model parameters at test time. Experimental results show the superiority of the proposed approach both in terms of perceptual quality and intelligibility of the enhanced speech signal.
\end{abstract}
\section{Introduction}
\label{sec:intro}

Speech enhancement is one of the central problems in audio signal processing \cite{loizou2007speech}. The goal is to recover a clean speech signal after observing a noisy mixture. In this work, we address single-channel speech enhancement, which can be seen as an under-determined source separation problem, where the sources are of different nature.

One popular statistical approach for source separation combines a local Gaussian model of the time-frequency signal coefficients with a variance model \cite{vincent2010probabilistic}. In this framework, non-negative matrix factorization (NMF) techniques have been used to model the time-frequency-dependent signal variance \cite{ISNMF,mohammadiha2013supervised}. Recently, discriminative approaches based on deep neural networks (DNNs) have also been investigated for speech enhancement, with the aim of estimating either clean spectrograms or time-frequency masks, given noisy spectrograms \cite{wang2017supervised, xu2015regression, weninger2015speech}. As a representative example, a DNN is used in \cite{xu2015regression} to map noisy speech log-power spectrograms into clean speech log-power spectrograms.

Even more recently, generative models based on deep learning, and in particular variational autoencoders (VAEs) \cite{kingma2013auto}, have been used for single-channel \cite{bando2017statistical,Leglaive_MLSP18} and multi-channel speech enhancement \cite{BayesianMVAE,Leglaive_ICASSP2019a}. These generative model-based approaches provide important advantages and justify the interest of semi-supervised methods for speech enhancement. Indeed, as shown in \cite{bando2017statistical, Leglaive_MLSP18}, fully-supervised methods such as \cite{xu2015regression} may have issues for generalizing to unseen noise types. The method proposed in \cite{Leglaive_MLSP18} was shown to outperform both a semi-supervised NMF baseline and the fully-supervised deep learning approach \cite{xu2015regression}.

In most cases, probabilistic models for source separation or speech enhancement rely on a Gaussianity assumption, which turns out to be restrictive for audio signals \cite{liutkus2015generalized}. As a result, \emph{heavy-tailed} distributions have started receiving increasing attention in the audio processing community \cite{simsekli2015alpha,liutkus2015cauchy,yoshii2016student}. In particular, $\alpha$-stable distributions (cf.\ Section~\ref{sec:model}) are becoming popular heavy-tailed models for audio modeling due to their nice theoretical properties \cite{kuruoglu1999signal,liutkus2015generalized,simsekli2015alpha,leglaive:hal-01416366,fontaine2017explaining,fontaine:lvaica_2018,simsekli2018alpha}.

In this work, we investigate the combination of a deep learning-based generative speech model with a heavy-tailed $\alpha$-stable noise model. The rationale for introducing a noise model based on heavy-tailed distributions as opposed to a structured NMF approach as in~\cite{Leglaive_MLSP18} is to avoid relying on restricting assumptions regarding stationarity or temporal redundancy of the noisy environment, that may be violated in practice, leading to errors in the estimates. In addition, we let the noise model remain unsupervised in order to avoid the aforementioned generalization issues regarding the noisy recording environment. We develop a Monte Carlo expectation-maximization algorithm \cite{wei1990monte} for performing maximum likelihood estimation at test time. Experiments performed under challenging conditions show that the proposed approach outperforms the competing approaches in terms of both perceptual quality and intelligibility. 


\section{Speech model}
\label{sec:speech_model}

We work in the short-term Fourier transform (STFT) domain where $\mbb{B}={\{0,...,F-1\}}\times{\{0,...,N-1\}}$ denotes the set of time-frequency bins. For  $(f,n) \in \mbb{B}$, $f$ denotes the frequency index and $n$ the time-frame index. We use $s_{fn}, b_{fn}, x_{fn} \in \mathbb{C}$ to denote the complex STFT coefficients of the speech, noise, and mixture signals, respectively. 

\begin{figure}[t]
\centering
\subfloat[Generative network.]{\includegraphics[width=.48\linewidth]{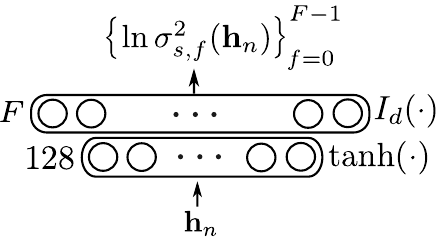}\label{fig:genNet}}
\hspace{0.2cm}
\subfloat[Recognition network.]{\includegraphics[width=.48\linewidth]{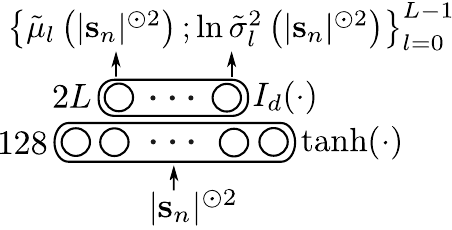}\label{fig:recNet}}
\caption{Generative and recognition networks. Beside each layer is indicated its size and the activation function.}
\label{fig:fullVAE}
\end{figure}

As in \cite{bando2017statistical,Leglaive_MLSP18}, independently for all $(f,n) \in \mathbb{B}$, we consider the following generative speech model involving a latent random vector $\mathbf{h}_n \in \mathbb{R}^L$, with $L \ll F$:
\begin{align}
\mathbf{h}_n &\sim \mathcal{N}(\mathbf{0}, \mathbf{I}); \label{prior_VAE} \\
s_{fn} \mid \mathbf{h}_n &\sim \mathcal{N}_c(0, \sigma_{s,f}^2(\mathbf{h}_n)),
\label{decoder_VAE}
\end{align}
where $\mathcal{N}(\mathbf{x} ; \bs{\mu}, \bs{\Sigma})$ denotes the multivariate Gaussian distribution for a real-valued random vector, $\mathbf{I}$ is the identity matrix of appropriate size, and $\mathcal{N}_c(x; \mu, \sigma^2)$ denotes the univariate complex proper Gaussian distribution. As represented in Fig.~\ref{fig:genNet}, $\{\sigma_{s,f}^2: \mbb{R}^L \mapsto \mbb{R}_+\}_{f=0}^{F-1}$ is a set of non-linear functions corresponding to a neural network which takes as input $\mbf{h}_n \in \mathbb{R}^L$. This variance term can be understood as a model for the short-term power spectral density of speech \cite{liutkus2011gaussian}. We denote by $\bs{\theta}_s$ the parameters of this \emph{generative neural network}. 

An important contribution of VAEs \cite{kingma2013auto} is to provide an efficient way of learning the parameters of such a generative model. Let $\mbf{s} = \{ \mbf{s}_n \in \mbb{C}^F \}_{n=0}^{N-1}$ be a training dataset of clean-speech STFT time frames and $\mbf{h} = \{ \mbf{h}_n \in \mbb{R}^L \}_{n=0}^{N-1}$ the set of associated latent random vectors. Taking ideas from variational inference, VAEs estimate the parameters $\bs{\theta}_s$ by maximizing a lower bound of the log-likelihood $\ln p(\mbf{s} ; \bs{\theta}_s)$ defined by:
\begin{equation}
\mathcal{L}\left(\bs{\theta}_s, \bs{\psi}\right) \hspace{-1.5pt}=\hspace{-1.5pt}  \E_{q\left(\mathbf{h} \mid \mathbf{s} ; \bs{\psi}\right)}\left[ \ln p\left(\mathbf{s} \mid \mathbf{h} ; \bs{\theta}_s \right) \right] - D_{\text{KL}}\left(q\left(\mathbf{h} \mid \mathbf{s} ; \bs{\psi}\right) \parallel p(\mathbf{h})\right),
\label{variational_bound}
\end{equation}
where $q\left(\mathbf{h} \mid \mathbf{s} ; \bs{\psi}\right)$ denotes an approximation of the intractable true posterior distribution $p(\mathbf{h} \mid \mathbf{s} ; \bs{\theta}_s )$, and $D_{\text{KL}}(q \parallel p) = \mathbb{E}_q[\ln(q/p)]$ is the Kullback-Leibler divergence. Independently for all the dimensions $l \in \{0,...,L-1\}$ and all the time frames $n \in \{0,...,N-1\}$, $q(\mathbf{h} \mid \mathbf{s} ; \bs{\psi})$ is defined by:
\begin{equation}
h_{l,n} \mid \mathbf{s}_n \sim \mathcal{N}\left(\tilde{\mu}_l\left(|\mathbf{s}_n|^{\odot 2}\right), \tilde{\sigma}_l^2\left(|\mathbf{s}_n|^{\odot 2}\right) \right),
\label{encoder_VAE}
\end{equation}
where $h_{l,n} = (\mathbf{h}_n)_l$ and $(\cdot)^{\odot \cdot}$ denotes element-wise exponentiation. As represented in Figure~\ref{fig:recNet}, $\{\tilde{\mu}_l: \mathbb{R}_+^{F} \mapsto \mathbb{R}\}_{l=0}^{L-1}$ and $\{\tilde{\sigma}_l^2: \mathbb{R}_+^{F} \mapsto \mathbb{R}_+\}_{l=0}^{L-1}$ are non-linear functions corresponding to a neural network which takes as input the speech power spectrum at a given time frame. $\bs{\psi}$ denotes the parameters of this \emph{recognition network}, which also have to be estimated by maximizing the \emph{variational lower bound} defined in \eqref{variational_bound}. Using \eqref{prior_VAE}, \eqref{decoder_VAE} and \eqref{encoder_VAE} we can develop this objective function as follows:
\begin{align}
\mathcal{L}\left(\bs{\theta}_s, \bs{\psi}\right) \overset{c}{=}&- \sum_{f=0}^{F-1}\sum_{n=0}^{N-1} \mathbb{E}_{q\left(\mathbf{h}_n \mid \mathbf{s}_n ; \bs{\psi} \right)}\left[ d_{\text{IS}}\left(\left|s_{fn}\right|^2 ; \sigma_f^2(\mathbf{h}_n)\right) \right] \nonumber \\
& \hspace{-1.6cm} + \frac{1}{2} \sum_{l=0}^{L-1} \sum_{n=0}^{N-1}\left[ \ln \tilde{\sigma}_l^2\left(\left|\mathbf{s}_n\right|^{\odot 2}\right) - \tilde{\mu}_l\left(\left|\mathbf{s}_n\right|^{\odot 2}\right)^2 - \tilde{\sigma}_l^2\left(\left|\mathbf{s}_n\right|^{\odot 2}\right) \right],
\label{ELBO}
\end{align}
where $d_{\text{IS}}(x;y) = x/y - \ln(x/y) - 1$ is the Itakura-Saito divergence. Finally, using the so-called ``reparametrization trick'' \cite{kingma2013auto} to approximate the intractable expectation in \eqref{ELBO}, we obtain an objective function which is differentiable with respect to both $\bs{\theta}_s$ and $\bs{\psi}$ and can be optimized using gradient-ascent-based algorithms. It is important to note that the only reason why the recognition network is introduced is to learn the parameters of the generative network.

\section{Noise and Mixture Models}
\label{sec:model}

In the previous section we have seen how to learn the parameters of the generative model \eqref{prior_VAE}-\eqref{decoder_VAE}. This model can then be used as a speech signal probabilistic prior for a variety of applications. In this paper we are interested in single-channel speech enhancement.

We do not assume prior knowledge about the recording environment, so that the noise model remains unsupervised. Independently for all $(f,n) \in \mathbb{B}$, the STFT coefficients of noise are modeled as complex circularly symmetric $\alpha$-stable random variables \cite{samorodnitsky1994stable}:
\begin{equation}
b_{fn} \sim \mathcal{S}\alpha\mathcal{S}(\sigma_{b,f}),
\label{noise_marginal_model}
\end{equation}
where $\alpha \in ]0,2]$ is the characteristic exponent and $\sigma_{b,f} \in \mathbb{R}_+$ is the scale parameter. As proposed in \cite{fontaine:lvaica_2018} for multichannel speech enhancement, this scale parameter is only frequency-dependent, it does not depend on time. For algorithmic purposes, the noise model \eqref{noise_marginal_model} can be conveniently rewritten in an \emph{equivalent} scale mixture of Gaussians form \cite{andrews1974scale}, by making use of the product property of the symmetric $\alpha$-stable distribution \cite{samorodnitsky1994stable}:
\begin{align}
\phi_{fn} &\sim \mathcal{P}\frac{\alpha}{2}\mathcal{S}\Big(2 \cos(\pi \alpha / 4)^{2/\alpha}\Big); \\
b_{fn} \mid \phi_{fn} &\sim \mathcal{N}_c(0, \phi_{fn} \sigma_{b,f}^2),
\label{noise_conditional_model}
\end{align}
where $\phi_{fn} \in \mathbb{R}_+$ is called the \emph{impulse variable}. It locally modulates the variance of the conditional distribution of $b_{fn}$ given in \eqref{noise_conditional_model}. $\mathcal{P}\frac{\alpha}{2}\mathcal{S}$ denotes a \textit{positive} stable distribution of characteristic exponent $\alpha/2$. It corresponds to a right-skewed heavy-tailed distribution defined for non-negative random variables \cite{magron2017}. These impulse variables can be understood as carrying uncertainty about the stationary noise assumption made in the marginal model \eqref{noise_marginal_model}, where the scale parameter does not depend on the time-frame index. 

The observed mixture signal is modeled as follows for all $(f,n) \in \mathbb{B}$:
\begin{equation}
x_{fn} = \sqrt{g_n} s_{fn} + b_{fn},
\label{mixture}
\end{equation}
where $g_n \in \mathbb{R}_+$ represents a frame-dependent but frequency-independent gain. The importance of this parameter was experimentally shown in \cite{Leglaive_MLSP18}. We further consider the conditional independence of the speech and noise STFT coefficients so that:
\begin{equation}
x_{fn} \mid \mathbf{h}_n, \phi_{fn} \sim \mathcal{N}_c\left(0, g_n \sigma_{s,f}^2(\mathbf{h}_n) + \phi_{fn} \sigma_{b,f}^2 \right).
\label{likelihood}
\end{equation}

\section{Inference}
\label{sec:inference}

Let $\bs{\theta}_u = \left\{\mathbf{g} = \{g_n \in \mathbb{R}_+ \}_{n=0}^{N-1}, \boldsymbol{\sigma}_b^2 = \{\sigma_{b,f}^2 \in \mathbb{R}_+ \}_{f=0}^{F-1} \right\}$ be the set of model parameters to be estimated. For maximum likelihood estimation, in this section we develop a Monte-Carlo expectation maximization (MCEM) algorithm \cite{wei1990monte}, which iteratively applies the so-called E- and M-steps until convergence, which we detail below. Remember that the speech generative model parameters $\boldsymbol{\theta}_s$ have been learned during a training phase (see Section~\ref{sec:speech_model}). We denote by $\mathbf{x} = \{x_{fn}\}_{(f,n) \in \mathbb{B}}$ the set of observed data while $\mathbf{z} = \left\{\mathbf{h}_n, \boldsymbol{\phi}_n = \{\phi_{fn}\}_{f=0}^{F-1} \right\}_{n=0}^{N-1}$ is the set of latent variables. We will also use $\mathbf{x}_n = \{x_{fn}\}_{f=0}^{F-1}$ and $\mathbf{z}_n = \left\{\mathbf{h}_n, \boldsymbol{\phi}_n \right\}$ to respectively denote the set of observed and latent variables at a given time frame $n$.

\myparagraph{Monte Carlo E-Step} Let $\bs{\theta}_u^\star$ be the current (or the initial) value of the model parameters. At the E-step of a standard expectation-maximization algorithm, we would compute the following conditional expectation of the complete-data log-likelihood $Q(\bs{\theta}_u ; \bs{\theta}_u^\star) = \mathbb{E}_{p(\mathbf{z} \mid \mathbf{x} ; \bs{\theta}_s, \bs{\theta}_u^\star)} \left[ \ln p(\mathbf{x}, \mathbf{z} ; \bs{\theta}_s, \bs{\theta}_u) \right]$. However, this expectation cannot be here computed analytically. We therefore approximate $Q(\bs{\theta}_u ; \bs{\theta}_u^\star)$ using an empirical average:
\begin{align}
\tilde{Q}(\bs{\theta}_u; \bs{\theta}_u^\star) \overset{c}{=}& - \frac{1}{R} \sum_{r=1}^{R} \sum_{(f,n) \in \mathbb{B}} \Big[ \ln\left( g_n \sigma_{s,f}^2\left(\mathbf{h}_n^{(r)}\right) + \phi_{fn}^{(r)} \sigma_{b,f}^2 \right) \nonumber \\
& + |x_{fn}|^2 \left(g_n \sigma_{s,f}^2\left(\mathbf{h}_n^{(r)}\right) + \phi_{fn}^{(r)} \sigma_{b,f}^2\right)^{-1} \Big],
\label{QFunction_approx}
\end{align}
where $\overset{c}{=}$ denotes equality up to an additive constant, and $\mathbf{z}_n^{(r)} = \big\{\mathbf{h}_n^{(r)}, \boldsymbol{\phi}_n^{(r)} = \big\{\phi_{fn}^{(r)}\big\}_{f=0}^{F-1}\big\}$, $r \in \{1,...,R\}$, is a sample drawn from the posterior $p(\mathbf{z}_n \mid \mathbf{x}_n ; \bs{\theta}_s, \bs{\theta}_u^\star)$ using a Markov Chain Monte Carlo (MCMC) method. This approach forms the basis of the MCEM algorithm \cite{wei1990monte}. Note that unlike the standard EM algorithm, it does not ensure an improvement in the likelihood at each iteration. Nevertheless, some convergence results in terms of stationary points of the likelihood can be obtained under suitable conditions \cite{chan1995monte}. 

In this work we use a (block) Gibbs sampling algorithm \cite{Robert:2005:MCS:1051451}. From an initialization $\mathbf{z}_n^{(0)}$, it consists in iteratively sampling from the so-called full conditionals. More precisely, at the $m$-th iteration of the algorithm and independently for all $n \in \{0,...,N-1\}$, we first sample $\mathbf{h}_n^{(m)} \sim p\big(\mathbf{h}_n \mid \mathbf{x}_n, \boldsymbol{\phi}_n^{(m-1)} ; \bs{\theta}_s, \bs{\theta}_u^\star\big)$. Then, we sample $\phi_{fn}^{(m)} \sim p\big(\phi_{fn} \mid x_{fn}, \mathbf{h}_n^{(m)} ; \bs{\theta}_s, \bs{\theta}_u^\star\big)$ independently for all $f \in \{0,...,F-1\}$. Those two full conditionals are unfortunately analytically intractable, but we can use one iteration of the Metropolis-Hastings algorithm to sample from them. This approach corresponds to the Metropolis-within-Gibbs sampling algorithm \cite[p. 393]{Robert:2005:MCS:1051451}. One iteration of this method is detailed in Algorithm~\ref{algo:metropolis-within-gibbs}. The proposal distributions for $\mathbf{h}_n$ and $\phi_{fn}$ are respectively given in lines \ref{line:proposal_h} and \ref{line:proposal_phi}. The two acceptance probabilities required in lines \ref{line:acc_prob_h} and \ref{line:acc_prob_phi} are computed as follows:
\begin{equation}
\alpha_{n}^{(h)} = \min\left(\hspace{-2.4pt} 1, \displaystyle \frac{p\left(\tilde{\mathbf{h}}_n\right) \prod\limits_{f=0}^{F-1} p\left(x_{fn} \mid \tilde{\mathbf{h}}_n, \phi_{fn}^{(m-1)} ; \bs{\theta}_s, \bs{\theta}_u^\star\right) }{p\left(\mathbf{h}_n^{(m-1)}\right) \prod\limits_{f=0}^{F-1} p\left(x_{fn} \mid \mathbf{h}_n^{(m-1)}, \phi_{fn}^{(m-1)} ; \bs{\theta}_s, \bs{\theta}_u^\star\right) } \hspace{-2.4pt}\right);
\label{acc_prob_h}
\end{equation}
\begin{equation}
\alpha_{n}^{(\phi)} = \min\left(1, \displaystyle \frac{p\big( x_{fn} \mid \mathbf{h}_n^{(m)}, \tilde{\phi}_{fn} ; \bs{\theta}_s, \bs{\theta}_u^\star\big)}{p\big( x_{fn} \mid \mathbf{h}_n^{(m)}, \phi_{fn}^{(m-1)} ; \bs{\theta}_s, \bs{\theta}_u^\star\big)}\right).
\label{acc_prob_phi}
\end{equation}
The two distributions involved in the computation of those acceptance probabilities are defined in \eqref{prior_VAE} and \eqref{likelihood}. Finally, we only keep the last $R$ samples for computing $\tilde{Q}(\bs{\theta}_u; \bs{\theta}_u^\star)$, i.e. we discard the samples drawn during a so-called burn-in period.

\myparagraph{M-Step} At the M-step we want to minimize $-\tilde{Q}(\bs{\theta}_u; \bs{\theta}_u^\star)$ with respect to $\bs{\theta}_u$. Let $\mathcal{C}(\bs{\theta}_u)$ denote the cost function associated with this non-convex optimization problem. Similar to \cite{fevotte2011algorithms}, we adopt a majorization-minimization approach. Let us introduce the two following sets of \emph{auxiliary variables} $\mathbf{c} = \{c_{fn}^{(r)} \in \mathbb{R}_+\}_{r,f,n}$ and $\boldsymbol{\lambda} = \{\lambda_{k,fn}^{(r)} \in \mathbb{R}_+\}_{k,r,f,n}$. We can show using standard concave/convex inequalities (see e.g. \cite{fevotte2011algorithms}) that $\mathcal{C}(\bs{\theta}_u) \le \mathcal{G}(\bs{\theta}_u, \mathbf{c}, \boldsymbol{\lambda})$, where:
\begin{align}
\mathcal{G}(\bs{\theta}_u, \mathbf{c}, \boldsymbol{\lambda}) =& \frac{1}{R} \sum_{r=1}^{R} \sum_{(f,n) \in \mathbb{B}} \Bigg[ \ln\left(c_{fn}^{(r)}\right) \nonumber \\
&+ \frac{1}{c_{fn}^{(r)}} \left( g_n \sigma_{s,f}^2\left(\mathbf{h}_n^{(r)}\right) + \phi_{fn}^{(r)} \sigma_{b,f}^2  - c_{fn}^{(r)}\right) \nonumber \\
&+ |x_{fn}|^2 \Bigg( \frac{\left(\lambda_{1,fn}^{(r)}\right)^2}{g_n \sigma_{s,f}^2\left(\mathbf{h}_n^{(r)}\right)} + \frac{\left(\lambda_{2,fn}^{(r)}\right)^2}{\phi_{fn}^{(r)} \sigma_{b,f}^2} \Bigg) \Bigg].
\end{align}
Moreover, this upper bound is tight, i.e. $\mathcal{C}(\bs{\theta}_u) = \mathcal{G}(\bs{\theta}_u, \mathbf{c}, \boldsymbol{\lambda})$, for
\begin{align}
c_{fn}^{(r)} &= g_n \sigma_{s,f}^2\left(\mathbf{h}_n^{(r)}\right) + \phi_{fn}^{(r)} \sigma_{b,f}^2; \label{aux_var_c}\\
\lambda_{1,fn}^{(r)} &= g_n \sigma_{s,f}^2\left(\mathbf{h}_n^{(r)}\right) \left(g_n \sigma_{s,f}^2\left(\mathbf{h}_n^{(r)}\right) + \phi_{fn}^{(r)} \sigma_{b,f}^2 \right)^{-1};\label{aux_var_lambda_1}\\
\lambda_{2,fn}^{(r)} &= \phi_{fn}^{(r)} \sigma_{b,f}^2 \left(g_n \sigma_{s,f}^2\left(\mathbf{h}_n^{(r)}\right) + \phi_{fn}^{(r)} \sigma_{b,f}^2\right)^{-1}.\label{aux_var_lambda_2}
\end{align}
Minimizing $\mathcal{G}(\bs{\theta}_u, \mathbf{c}, \boldsymbol{\lambda})$ with respect to the model parameters $\bs{\theta}_u$ is a convex optimization problem. By zeroing the partial derivatives of $\mathcal{G}$ with respect to each scalar in $\bs{\theta}_u$ we obtain update rules that depend on the auxiliary variables. We then replace these auxiliary variables with the formulas given in \eqref{aux_var_c}-\eqref{aux_var_lambda_2}, which makes the upper bound $\mathcal{G}$ tight.
This procedure ensures that the cost $\mathcal{C}(\bs{\theta}_u)$ decreases \cite{hunter2004tutorial}. The resulting final updates are given as follows:
\begin{align}
\sigma_{b,f}^2 &\leftarrow \sigma_{b,f}^2 \left[ \frac{\sum\limits_{n=0}^{N-1} |x_{fn}|^2  \sum\limits_{r=1}^{R} \phi_{fn}^{(r)} \left(v_{x,fn}^{(r)}\right)^{-2}}{\sum\limits_{n=0}^{N-1} \sum\limits_{r=1}^{R} \phi_{fn}^{(r)} \left(v_{x,fn}^{(r)}\right)^{-1}} \right]^{1/2}; \\
g_n &\leftarrow g_n \left[ \frac{\sum\limits_{f=0}^{F-1} |x_{fn}|^2  \sum\limits_{r=1}^{R} \sigma_{s,f}^2\left(\mathbf{h}_n^{(r)}\right) \left(v_{x,fn}^{(r)}\right)^{-2}}{\sum\limits_{f=0}^{F-1} \sum\limits_{r=1}^{R} \sigma_{s,f}^2\left(\mathbf{h}_n^{(r)}\right) \left(v_{x,fn}^{(r)}\right)^{-1}} \right]^{1/2},
\end{align}
where we introduced $v_{x,fn}^{(r)} = g_n \sigma_{s,f}^2\left(\mathbf{h}_n^{(r)}\right) + \phi_{fn}^{(r)}\sigma_{b,f}^2$ in order to ease the notations. Non-negativity is ensured provided that the parameters are initialized with non-negative values. 

\myparagraph{Speech reconstruction} Once the unsupervised model parameters $\bs{\theta}_u$ are estimated with the MCEM algorithm, we need to estimate the clean speech signal. For all $(f,n) \in \mathbb{B}$, let $\tilde{s}_{fn} = \sqrt{g_n}s_{fn}$ be the scaled version of the speech STFT coefficients. We estimate these variables according to their posterior mean, given by:
\begin{align}
\hat{\tilde{s}}_{fn} &= \mathbb{E}_{p(\tilde{s}_{fn} \mid x_{fn} ; \bs{\theta}_u, \bs{\theta}_s)}  [\tilde{s}_{fn}] \nonumber \\
&= \mathbb{E}_{p(\mathbf{z}_n \mid \mathbf{x}_n ; \bs{\theta}_u, \bs{\theta}_s) }\left[ \mathbb{E}_{p(\tilde{s}_{fn} \mid \mathbf{z}_n, \mathbf{x}_n ; \bs{\theta}_u, \bs{\theta}_s)} [\tilde{s}_{fn}] \right] \nonumber \\
&= \mathbb{E}_{p(\mathbf{z}_n \mid \mathbf{x}_n ; \bs{\theta}_u, \bs{\theta}_s) }\left[ \frac{g_n \sigma_{s,f}^2(\mathbf{h}_n)}{g_n \sigma_{s,f}^2(\mathbf{h}_n) + \phi_{fn} \sigma_{b,f}^2} \right] x_{fn}.
\end{align} 
As before, this expectation cannot be computed analytically so it is approximated using the Metropolis-within-Gibbs sampling algorithm detailed in Algorithm~\ref{algo:metropolis-within-gibbs}. This estimate corresponds to a probabilistic Wiener filtering averaged over all possible realizations of the latent variables according to their posterior distribution.

\begin{algorithm}[t]
\caption{$m$-th iteration of the Metropolis-within-Gibbs sampling algorithm}\label{MCE-Step}
\begin{algorithmic}[1]
\ForAll{$n \in \{0,...,N-1\}$}
\State Sample $\tilde{\mathbf{h}}_n \sim \mathcal{N}(\mathbf{h}_n^{(m-1)}, \epsilon^2 \mathbf{I})$ \label{line:proposal_h}
\State Compute acceptance probability $\alpha_{n}^{(h)}$ (equation \eqref{acc_prob_h}) \label{line:acc_prob_h}
\State Set $\mathbf{h}_n^{(m)} = \begin{cases}
\tilde{\mathbf{h}}_n & \text{ if } \alpha_{n}^{(h)} > u_{n}^{(h)} \sim \mathcal{U}([0,1]) \\
\mathbf{h}_n^{(m-1)} & \text{ otherwise}
\end{cases}$
\ForAll{$f \in \{0,...,F-1\}$}
\State Sample $\tilde{\phi}_{fn} \sim \mathcal{P}\frac{\alpha}{2}\mathcal{S}\Big(2 \cos(\pi \alpha / 4)^{2/\alpha}\Big)$ \label{line:proposal_phi}
\State Compute acceptance probability $\alpha_{n}^{(\phi)}$ (equation \eqref{acc_prob_phi}) \label{line:acc_prob_phi}
\State Set $\phi_{fn}^{(m)} = \begin{cases}
\tilde{\phi}_{fn} & \text{ if } \alpha_{n}^{(\phi)} > u_n^{(\phi)} \sim \mathcal{U}([0,1]) \\
\phi_{fn}^{(m-1)} & \text{ otherwise}
\end{cases}$
\EndFor
\EndFor
\end{algorithmic}
\vspace{-2pt}
\label{algo:metropolis-within-gibbs}
\end{algorithm}

\vspace{-5pt}
\section{Experiments}
\label{sec:experiments}

\begin{figure*}[ht]
\centering
\includegraphics[width=.95\linewidth]{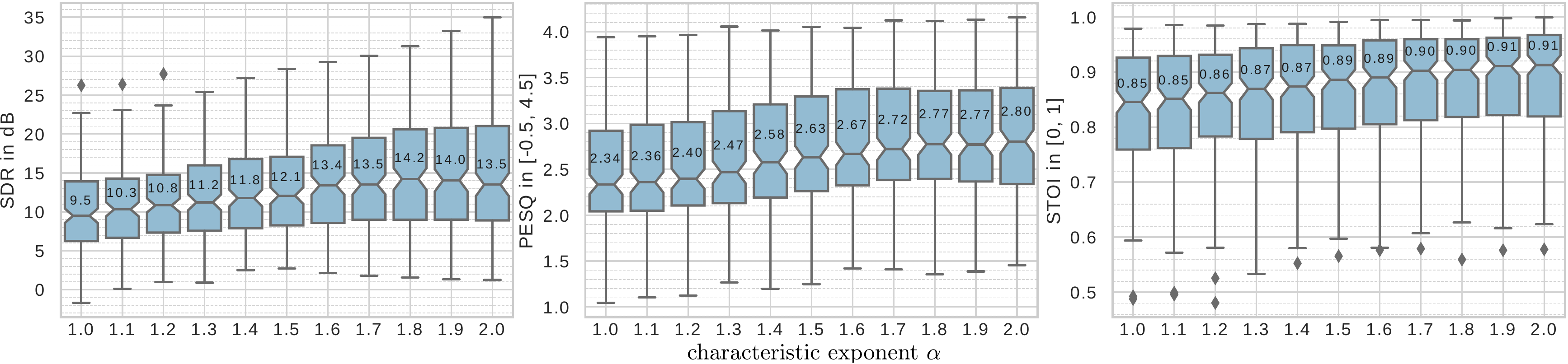}
\vspace{-5pt}
\caption{Speech enhancement results obtained with the proposed method as a function of the characteristic exponent $\alpha$ in the noise model \eqref{noise_marginal_model}. $\alpha=2.0$ actually corresponds to $\alpha=1.999$. The value of the median is indicated within each boxplot.}
\label{fig:results_vs_alpha}
\vspace{-10pt}
\end{figure*}

\begin{figure}[ht]
\centering
\includegraphics[width=\linewidth]{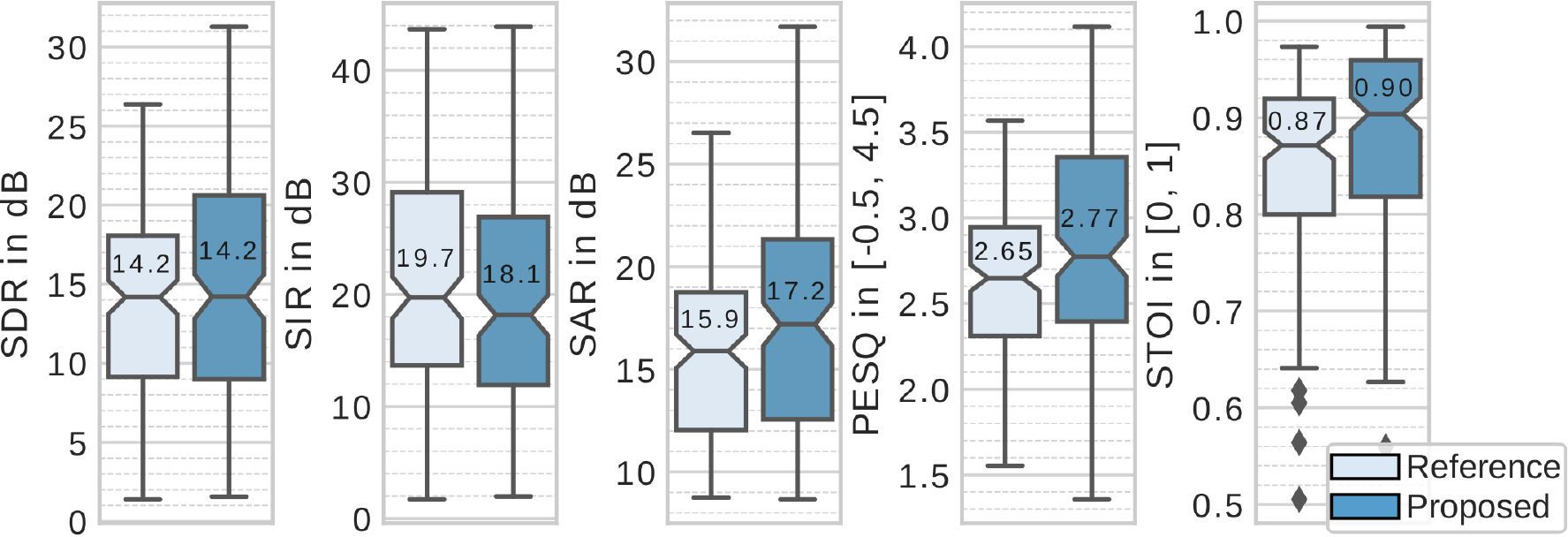}
\vspace{-5pt}
\caption{Results obtained with the reference and the proposed methods (using $\alpha=1.8$). The median is indicated within each boxplot.}
\label{fig:final_results}
\vspace{-10pt}
\end{figure}

\textit{Reference method}: The proposed approach is compared with the recent method \cite{Leglaive_MLSP18}. The speech signal in this paper is modeled in the exact same manner as in the current work, only the noise model differs. This latter is a Gaussian model with an NMF parametrization of the variance \cite{ISNMF}: $b_{fn} \sim \mathcal{N}_c\left(0, \left(\mathbf{W}_b\mathbf{H}_b\right)_{f,n}\right)$, where $\mathbf{W}_b \in \mathbb{R}_+^{F \times K}$ and  $\mathbf{H}_b \in \mathbb{R}_+^{K \times N}$. It is also unsupervised in the sense that both $\mathbf{W}_b$ and $\mathbf{H}_b$ are estimated from the noisy mixture signal. This method also relies on an MCEM algorithm. By comparing the proposed method with \cite{Leglaive_MLSP18}, we fairly investigate which noise model leads to the best speech enhancement results. Note that the method proposed in \cite{Leglaive_MLSP18} was shown to outperform both a semi-supervised NMF baseline and the fully-supervised deep learning approach \cite{xu2015regression}, the latter having difficulties for generalizing to unseen noise types. We do not include here the results obtained with these two other methods. 

\textit{Database}: The supervised speech model parameters in the proposed and the reference methods are learned from the training set of the TIMIT database \cite{TIMIT}. It contains almost 4 hours of 16-kHz speech signals, distributed over 462 speakers. For the evaluation of the speech enhancement algorithms, we mixed clean speech signals from the TIMIT test set and noise signals from the DEMAND database \cite{DEMAND_db}, corresponding to various noisy environments: domestic, nature, office, indoor public spaces, street and transportation. We created 168 mixtures at a 0~dB signal-to-noise ratio (one mixture per speaker in the TIMIT test set). Note that both speakers and sentences are different than in the training set.

\textit{Parameter setting}: The STFT is computed using a 64-ms sine window (i.e.~$F=513$) with 75\%-overlap. Based on \cite{Leglaive_MLSP18}, the latent dimension in the speech generative model \eqref{prior_VAE}-\eqref{decoder_VAE} is fixed to $L=64$. In this reference method, the NMF rank of the noise model is fixed to $K=10$. The NMF parameters are randomly initialized. For both the proposed and the reference method, the gain $g_n$ is initialized to one for all time frames $n$. For the proposed method, the noise scale parameter $\sigma_{b,f}$ is also initialized to one for all frequency bins. We run 200 iterations of the MCEM algorithm. At each Monte-Carlo E-Step, we run 40 iterations of the Metropolis-within-Gibbs algorithm and we discard the first 30 samples as the burn-in period. The parameter $\epsilon^2$ in line~\ref{line:proposal_h} of Algorithm~\ref{algo:metropolis-within-gibbs} is set to $0.01$.

\textit{Neural network}: The structure of the generative and recognition networks is the same as in \cite{Leglaive_MLSP18} and is represented in Fig.~\ref{fig:fullVAE}. Hidden layers use hyperbolic tangent ($\tanh(\cdot)$) activation functions and output layers use identity activation functions ($I_d(\cdot)$). The output of these last layers is therefore real-valued, which is the reason why we consider logarithm of variances.  For learning the parameters $\bs{\theta}_s$ and $\bs{\phi}$, we use the Adam optimizer \cite{kingma2014adam} with a step size of $10^{-3}$, exponential decay rates for the first and second moment estimates of $0.9$ and $0.999$ respectively, and an epsilon of $10^{-7}$ for preventing division by zero. 20\% of the TIMIT training set is kept as a validation set, and early stopping with a patience of 10 epochs is used. Weights are initialized using the uniform initializer described in \cite{glorot2010understanding}.

\textit{Results}: The estimated speech signal quality is evaluated in terms of standard energy ratios expressed in decibels (dBs) \cite{vincent2006performance}: the signal-to-distortion (SDR), signal-to-interference (SIR) and signal-to-artifact (SAR) ratios. We also consider the perceptual evaluation of speech quality (PESQ) \cite{rix2001perceptual} measure (in $[-0.5, 4.5]$), and the short-time objective intelligibility (STOI) measure \cite{taal2011algorithm} (in $[0, 1]$). For all measures, the higher the better. We first study the performance of the proposed method according to the choice of the characteristic exponent $\alpha$ in the noise model \eqref{noise_marginal_model}. Results presented in Fig.~\ref{fig:results_vs_alpha} indicate that according to the PESQ and STOI measures, the best performance is obtained for $\alpha=2$ (Gaussian case).\footnote{Actually $\alpha=1.999$ because for $\alpha=2$ the positive $\alpha$-stable distribution in \eqref{noise_marginal_model} is degenerate.} The SDR indicates that we should choose $\alpha=1.8$. Indeed, for greater values of $\alpha$, the SIR starts to decrease ($\sim \hspace{-3pt} 1$dB difference between $\alpha =1.8$ and $\alpha =2$) while the SAR remains stable (results are not shown here due to space constraints). Therefore, in Fig.~\ref{fig:final_results} we compare the results obtained with the reference \cite{Leglaive_MLSP18} and the proposed method using $\alpha=1.8$. With the proposed method, the estimated speech signal contains more interferences (SIR is lower) but less artifacts (SAR is higher). According to the SDR both methods are equivalent. But for intelligibility and perceptual quality, artifacts are actually more disturbing than interferences \cite{venkataramani2018performance}, which is the reason why the proposed method obtains better results in terms of both STOI and PESQ measures. For reproducibility, a Python implementation of our algorithm and audio examples are available online.\footnote{\url{https://team.inria.fr/perception/icassp2019-asvae/}}

\vspace{-2pt}
\section{Conclusion}
\label{sec:conclusion}
\vspace{-2pt}

In this work, we proposed a speech enhancement method exploiting a speech model based on VAEs and a noise model based on alpha-stable distributions. At the expense of more interferences, the proposed $\alpha$-stable noise model reduces the amount of artifacts in the estimated speech signal, compared to the use of a Gaussian NMF-based noise model as in \cite{Leglaive_MLSP18}. Overall, it is shown that the proposed approach improves the intelligibility and perceptual quality of the enhanced speech signal. Future works include extending the proposed approach to a multi-microphone setting using multivariate $\alpha$-stable distributions \cite{leglaive:hal-01416366, fontaine:lvaica_2018}.

\bibliographystyle{myIEEEbib}
\balance
\bibliography{IEEEabrv,refs}

\end{document}